\documentclass[preprint,12pt]{elsarticle}

\usepackage{amssymb}

\begin{document}

\begin{frontmatter}



\title{Quantum entanglement and phase transition in a two-dimensional photon-photon pair model}


\author{Jian-Jun Zhang}
\ead{ruoshui789@gmail.com}
\author{Jian-Hui Yuan}
\author{Jun-Pei zhang}
\author{Ze Cheng}

\address{School of Physics, Huazhong University of Science and
Technology,  Wuhan 430074, China}

\begin{abstract}
We propose a two-dimensional model consisting of photons and photon pairs. In the model, the mixed gas of photons and photon pairs is formally equivalent to a two-dimensional system of massive bosons with non-vanishing chemical potential, which implies the existence of two possible condensate phases. Using the variational method, we discuss the quantum phase transition of the mixed gas and obtain the critical coupling line analytically. Moreover, we also find that the phase transition of the photon gas can be interpreted as second harmonic generation. We then discuss the entanglement between photons and photon pairs. Additionally, we also illustrate how the entanglement between photons and photon pairs can be associated with the phase transition of the system.

\end{abstract}


\end{frontmatter}


\section{Introduction}
Bose-Einstein condensate (BEC) is the remarkable state of matter that spontaneously emerges when a system of bosons becomes cold enough that a significant fraction of them condenses into a single quantum state to minimize the system's free energy. Particles in that state then act collectively as a coherent wave. The phase transition for an atomic gas was first predicted by Einstein in 1924 and experimentally confirmed with the discovery of superfluid helium-4 in 1938. Obviously, atoms aren't the only option for a BEC. In recent years, with the development of techniques, the phenomenon of BEC was observed in several physical system [1-9], including exciton polaritions, solid-state quasiparticles and so on. We know that photons are the simplest of bosons, so that it would seem that they could in principle undergo this kind of condensation. The difficulty is that in the usual blackbody configuration, which consists of an empty three-dimensional (3D) cavity, the photon is massless and its chemical potential is zero, so that the BEC of photons under these circumstances would seem to be impossible. However, very recently, J. Klaers, etc. have overcome both obstacles using a simple approach [10,11]: By confining laser light within a two-dimensional (2D) cavity bounded by two concave mirrors, they create the conditions required for light to thermally equilibrate as a gas of conserved particles rather than as ordinary blackbody radiation.

  What is more, it is well known that there are many fascinated optical effects in the nonlinear medium, for instance, reduced fluctuation in one quadrature (squeezing) [12], sub-Poissonian statistics of the radiation field [13], or the collapse-revivals phenomenon [14]. Especially, in the nonlinear medium, a photon from the laser beam can couple with other photons to form a photon-pair (PP) [15-18]. The essence of PP has been investigated by many authors [19-21]. However, inspired by the experimental discovery of BEC of photons, in this letter we construct another interesting 2D model consisting of photons and PPs. In this model, the mixed system of photons and PPs is formally equivalent to a 2D gas of massive bosons with non-vanishing chemical potential, which implies the existence of two possible condensate phases, the mixed photon-PP condensate phase and the pure PP condensate phase. By means of a variational method we investigate the quantum phase transition of the mixed photon gas. Especially, we find that the quantum phase transition of the photon gas can be interpreted as second harmonic generation. We then discuss the entanglement between photons and PPs. By investigating the entanglement in the ground state and the dynamics of entanglement, we also illustrate how the entanglement between photons and PPs can be associated with the phase transition of the system. The investigation of these questions is important both for its connection with quantum optics and for its practical applications to harmonic generation and quantum information.

     The remainder of this paper is organized as follows: In Sec. II, we theoretically investigate the phenomenon of BEC of photons and PPs in a 2D optical microcavity. The entanglement between photons and PPs is investigated in Sec. III. Finally, we make a simple conclusion.

\section{Bose-Einstein condensation of photons and photon pairs}
\begin{figure}[b]
  \includegraphics[width=11cm,]{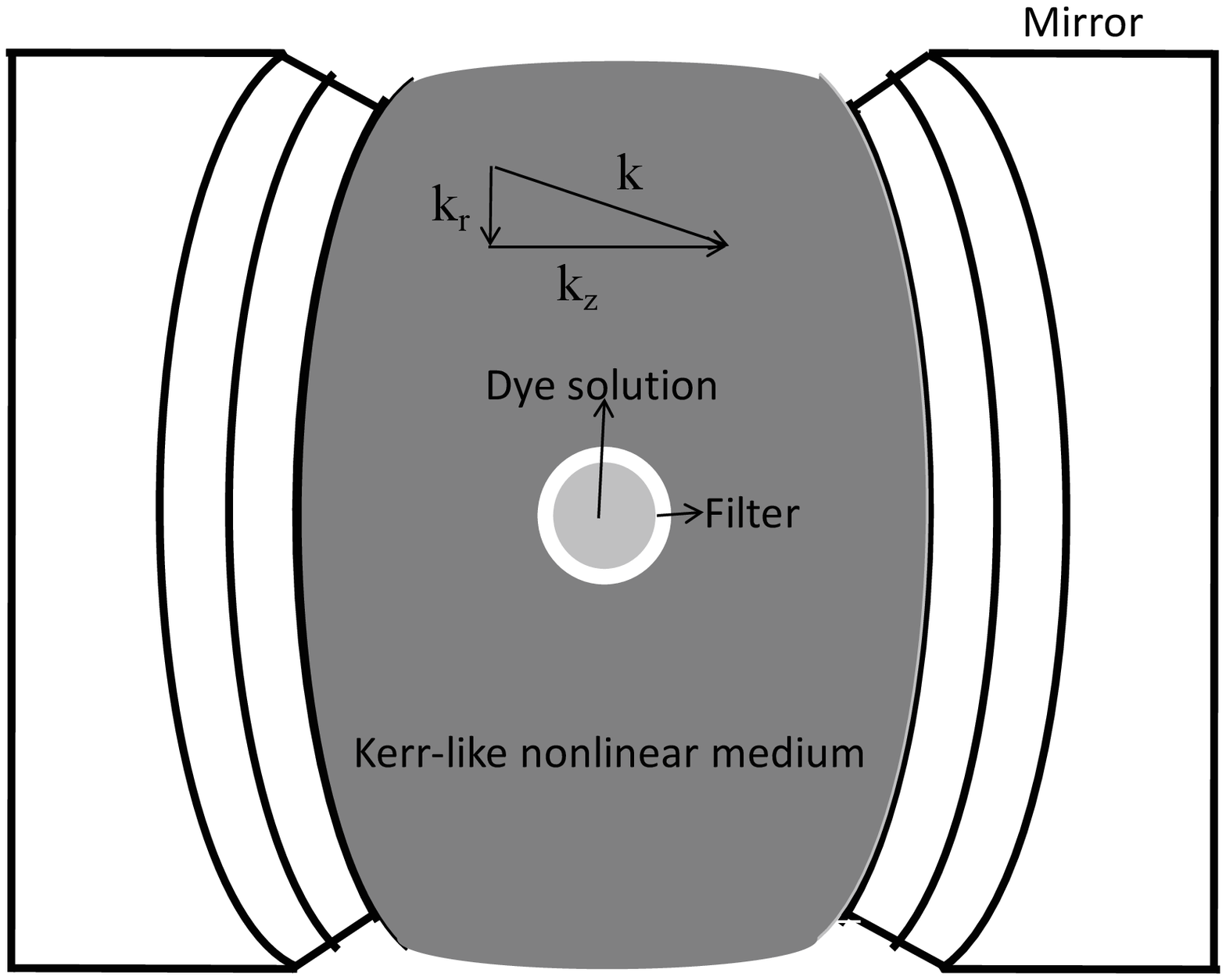}
  \caption{Scheme of the optical microcavity:
The microcavity consists of two curved-mirrors with high reflectivity. A filter filled with a dye solution is inserted into it, in which photons are repeatedly absorbed and re-emitted by the dye molecules. The other part of the cavity is filled with a Kerr-like nonlinear medium. }\label{Graph1.eps}
\end{figure}

\subsection {Description of the photon pair}

We start with the description of the PP. History speaking, the essence of the PP is presently still under discussion [19-21], and there exist many different ways to obtain it. However, here we will use the standard procedure [22] in the construction of harmonic generation to derive the PP. We know that the presence of an electromagnetic field in the nonlinear material causes a polarization of the medium and the polarization can be expanded in powers of the instantaneous electric field:
\begin{eqnarray}
{\bf{P}}({\bf{r}},t) = \chi ^{(1)} {\bf{E}}({\bf{r}},t) + \chi ^{(2)} {\bf{E}}^2 ({\bf{r}},t) + ...   .
\end{eqnarray}
Here, the first term defines the usual linear susceptibility, and the second term defines the lowest order nonlinear susceptibility. Ignoring the high order parts (i.e. only expend the polarization to second order in electric field $E$), we find that the Hamiltonian describing the interaction of the radiation field with the dielectric medium is decomposed into two terms:
\begin{eqnarray}
\begin{array}{l}
 {\kern 1pt} {\kern 1pt} {\kern 1pt} {\kern 1pt} {\kern 1pt} {\kern 1pt} {\kern 1pt} {\kern 1pt} {\kern 1pt} {\kern 1pt} {\kern 1pt} {\kern 1pt} {\kern 1pt} {\kern 1pt} H_{{\mathop{\rm int}} }  = H_{line}  + H_{nonline}  \\
 {\kern 1pt} {\kern 1pt} {\kern 1pt} {\kern 1pt} {\kern 1pt} H_{line}  =  - \int {\chi ^{(1)} {\bf{E}}^2 ({\bf{r}},t)} d{\bf{r}} \\
 H_{nonline}  =  - \int {\chi ^{(2)} {\bf{E}}^3 ({\bf{r}},t)} d{\bf{r}} \\.
 \end{array}
\end{eqnarray}
where $H_{line}$ represents the energy of the linear interaction and $H_{nonline}$ the nonlinear interaction.

It is well known that the electric field operator in a microcavity can be expanded in terms of normal modes [23] as
\begin{eqnarray}
{\bf{E}}(r,t) = i\sum\limits_{\bf{k}} {{\bf{e}}_{\bf{k}} } \left( {\frac{{\hbar \omega _{\bf{k}} }}{{2V\varepsilon }}} \right)^{1/2} \left( {\hat a_{\bf{k}} e^{ - i\omega _{\bf{k}} t + i{\bf{k}} \cdot {\bf{r}}}  + \hat a_{\bf{k}}^ +  e^{i\omega _{\bf{k}} t - i{\bf{k}} \cdot {\bf{r}}} } \right),
\end{eqnarray}
where $\hat a_{\bf{k}}$ and $\hat a_{\bf{k}}^ +$ are the annihilation and creation operators of photons with frequency $\omega _{\bf{k}}$, and they all obey the usual boson commutation rules. $V$ is the normalization volume, $\varepsilon$ is the dielectric constant of the medium and ${\bf{e}}_{\bf{k}}$ is the unit polarization vector with the usual polarization indices omitted for simplicity. Substituting (3) into (2), for the linear interaction part, we find that it consists of two processes, dissipation and two-photon absorption(or emission). Here, dissipation is essentially also a two-photon process, in which one photon is absorbed by the medium, meanwhile another one is emitted. The linear interaction can be ignored, if the incident photon field frequency $\omega _0$ is well below the electronic transition frequencies of the medium. In that case, we need only consider the nonlinear interaction, which has the simple form
\begin{eqnarray}
H_{nonline}  = \frac{\hbar }{{\sqrt V }}\sum\limits_{{\bf{k,k'}}} {\chi _{{\bf{k}},{\bf{k'}}} \left( {\hat b_{{\bf{k}} + {\bf{k'}}}^ +  \hat a_{\bf{k}} \hat a_{{\bf{k'}}}  + H.c.} \right)},
\end{eqnarray}
under the requirements of phase matching. Above, the operator $\hat a$ represents the normal photons, $\hat b$ represents the coupling PP, and where $\chi _{{\bf{k}},{\bf{k'}}}$ is the coupling matrix element. The interaction energy in (4) consists of two terms. The first term $b_{{\bf{k}} + {\bf{k'}}}^ +  a_{\bf{k}} a_{{\bf{k'}}}$ describes the process in which two normal photon with wave-vector ${\bf{k}}$ and ${\bf{k'}}$ couple into a PP with wave-vector ${\bf{K}} = {\bf{k}} + {\bf{k'}}$, and the second term describe the opposite process. The energy is conserved in both the processes.

\subsection {Free-photon dispersion relation inside the optical microcavity}
In this letter, we restrict out investigation inside a 2D optical microcavity. The microcavity, as shown in Fig. 1, consists of two curved dielectric mirrors with high reflectivity (about 99.9), which ensure prefect reflection of the longitudinal component of the electromagnetic field within the cavity. In addition, the transverse size of the cavity is much larger than its longitudinal one.

We know that for a free photon, its frequency as a function of transversal ($k_r$) and longitudinal ($k_z$) wave number is $\omega  = c\left[ {k_z^2  + k_r^2 } \right]^{1/2}$. However, in the case of photons confined inside the microcavity, the vanishing of the electric field at the reflecting surfaces of the curved-mirrors imposes a quantization condition on the longitudinal mode number $k_z$, $k_z  = n\pi /D(r)$, where $n$ is an integer and where $D(r) = D_0  - 2(R - \sqrt {R^2  - r^2 } )$ is the separation of two curved-mirrors at distance $r$ from the optical axis, with $D_0$ the mirror separation at distance $r = 0$ and $R$ the radius of curvature.

In the present work, we consider to fix the longitudinal mode number of photons by inserting a circular filter into the cavity. The filter is filled with a dye solution, in which photons are repeatedly absorbed and re-emitted by the dye molecules. Thus, it also plays the role of photon reservoir. We know that the longitudinal size of the cavity (i.e., the distance between the mirrors) is very small. The small distance $D(r)$ between the mirrors causes a large frequency spacing between adjacent longitudinal modes, comparable with the spectral width of the dye. Modify spontaneous emission such that the emission of photons with a given longitudinal mode number, $n = q$ in our case, dominates over other emission processes. In this way, the longitudinal mode number is frozen out. For fixed longitudinal mode number $q$ and in paraxial approximation ($r \ll R$, $k_r  \ll k_z$), we also find that the dispersion relation of photons approximatively becomes $\omega  \approx q\pi c/D_0  + ck_r^2 D_0 /2q\pi$. The above frequency-wavevector relation, upon multiplication by $\hbar$, becomes the energy-momentum relation for the photon
\begin{eqnarray}
E \approx m_{{\rm{ph}}} c^2  + \frac{{(p_r )^2 }}{{2m_{{\rm{ph}}} }},
\end{eqnarray}
where $m_{{\rm{ph}}}  = \hbar q\pi /D_0 c = \hbar \omega _{eff} /c^2$ is the effective mass of the confined photons. At low temperatures, it is convenient to redefine the zero of energy, so that only the effective kinetic energy,
\begin{eqnarray}
E \approx \frac{{(p_r )^2 }}{{2m_{{\rm{ph}}} }},
\end{eqnarray}
remains. The above analysis shows that for the photon confined inside the 2D microcavity, it is formally equivalent to a general boson having an effective mass $m_{{\rm{ph}}}  = \hbar \omega _{eff} /c^2$, that is moving in the transverse resonator plane.

Furthermore, we here consider the case that the microcavity (except the filter part) is filled with a Kerr nonlinear medium exhibiting significant third-order optical nonlinearity. Due to the nonlinear effect, photons can couple into PPs. If we connect the non-vanishing effective photon mass to the previous analysis of the PPs, in this case we then can rewrite the nonlinear interaction $H_{nonline}$ as
\begin{eqnarray}
H_{nonline}  = \frac{\hbar }{{\sqrt S }}\sum\limits_{{\bf{k}}_r {\bf{,k'}}_r } {\chi _{{\bf{k}}_r ,{\bf{k'}}_r } \left( {b_{{\bf{k}}_r  + {\bf{k'}}_r }^ +  a_{{\bf{k}}_r } a_{{\bf{k'}}_r }  + H.c.} \right)}
\end{eqnarray}
where $S$ is the surface area of the 2D cavity, and where $a_{{\bf{k}}_r }$ and $a_{{\bf{k'}}_r }$ are the annihilation operators of massive photons with transverse wavevectors ${\bf{k}}_r$ and ${\bf{k}}_{r'}$, respectively, and $b_{{\bf{k}}_r  + {\bf{k'}}_r }^ +$ are the creation operator of the massive PPs with transverse wave-vector $K_r  = {\bf{k}}_r  + {\bf{k'}}_r$. Here, it should be remarked that the existence of effective photon mass makes the thermodynamics of this 2D mixed gas of photons and PPs different from the usual 3D photon gas. For the 2D system, thermalization is achieved in a photon-number-conserving way ($N = N_a  + 2N_b $) with nonvanishing chemical potential $\mu$, by multiple scattering with the dye molecules, which acts as heat bath and equilibrates the transverse modal degrees of freedom of the photon gas to the temperature of dye molecules.

\subsection {BEC of photons and photon pairs}

In virtue of the above analysis, we consider the following basic Hamiltonian, to give a simple model of PP formation (with $\hbar  = 1$ throughout this letter)
\begin{eqnarray}
H_\mu   = H_a  + H_b  + H_{ab},
\end{eqnarray}
with
\begin{eqnarray}
\begin{array}{l}
 H_a  = \sum\limits_{\bf{k}} {\frac{{{\bf{k}}^2 }}{{2m_{ph} }}a_{\bf{k}}^ +  a_{\bf{k}}^{} }  + \frac{{u_{aa} }}{S}\sum\limits_{{\bf{k}},{\bf{k'}},{\bf{k''}}} {a_{{\bf{k}} + {\bf{k'}} - {\bf{k''}}}^ +  a_{{\bf{k''}}}^ +  } a_{{\bf{k'}}} a_{\bf{k}}  \\
 H_b  = \sum\limits_{\bf{k}} {\left( {\frac{{{\bf{k}}^2 }}{{4m_{ph} }} - 2\mu } \right)b_{\bf{k}}^ +  b_{\bf{k}}^{} }  + \frac{{u_{bb} }}{S}\sum\limits_{{\bf{k}},{\bf{k'}},{\bf{k''}}} {b_{{\bf{k}} + {\bf{k'}} - {\bf{k''}}}^ +  b_{{\bf{k''}}}^ +  } b_{{\bf{k'}}} b_{\bf{k}}  \\
 H_{ab}  = \frac{{u_{a,b} }}{S}\sum\limits_{{\bf{k}},{\bf{k'}}} {a_{\bf{k}}^ +  b_{{\bf{k'}}}^ +  } b_{{\bf{k'}}} a_{\bf{k}}  - \frac{1}{{\sqrt S }}\sum\limits_{{\bf{k}},{\bf{k'}}} {\chi _{{\bf{k}},{\bf{k'}}} (b_{{\bf{k}} + {\bf{k'}}}^ +  } a_{{\bf{k'}}} a_{\bf{k}}  + H.c.) \\
 \end{array}
\end{eqnarray}
Above, $H_a$ and $H_b$ denote the pure photon and PP contributions, and $H_{ab}$ refers to the interaction between them. In the dilute gas limit $u_{a,a}$, $u_{b,b}$, $u_{a,b}$ are proportional to the two-body s-wave photon-photon, photon-PP, and PP-PP scattering lengths [23], respectively, and $\chi _{{\bf{k}},{\bf{k'}}}$ characterizes the coupling strength, encoding that PPs are composed of two massive photons. Note that in (9) we ignore the chemical potential term $\mu N_a$ from the Hamiltonian [24]. Additionally, in (9), we also drop the subscript of the transverse wave-vectors of photons and PPs, i.e. we take ${\bf{k}}_r  = {\bf{k}}$.

It is well known that for a general massive Boson-Boson pairs gas, at absolute zero temperature, there exists a BEC consisting of two possible condensate phases [25-27]: (i) Both the single boson and the pair of bosons are condensed. (ii) The pair of bosons are condensed but the single boson is not. Now we know that in the case of photons (or PPs) confined inside the microcavity, the subsystem of photons (or PPs) is formally equivalent to a 2D gas of massive bosons with non-vanishing chemical potential. Thus, the feature should also survive for the 2D mixed gas of photons and PPs. In the condensate phase, a macroscopic number of particles occupy the zero-momentum state, and it is useful to separate out the condensate modes from the Hamiltonian. Follow the process, we find that at the BEC state, the grand canonical Hamiltonian of the mixed system has the form
\begin{eqnarray}
H_\mu   = H_0  + \delta H,
\end{eqnarray}
where
\begin{eqnarray}
\begin{array}{l}
 H_0  =  - \mu _b b^ +  b + \frac{{u_{aa} }}{S}\left( {a^ +  a} \right)^2  + \frac{{u_{bb} }}{S}\left( {b^ +  b} \right)^2  \\
 {\kern 1pt} {\kern 1pt} {\kern 1pt} {\kern 1pt} {\kern 1pt} {\kern 1pt} {\kern 1pt} {\kern 1pt} {\kern 1pt} {\kern 1pt} {\kern 1pt} {\kern 1pt} {\kern 1pt} {\kern 1pt} {\kern 1pt} {\kern 1pt} {\kern 1pt} {\kern 1pt} {\kern 1pt} {\kern 1pt} {\kern 1pt} {\kern 1pt} {\kern 1pt} {\kern 1pt} {\kern 1pt} {\kern 1pt} {\kern 1pt} {\kern 1pt}  + \frac{{u_{ab} }}{S}a^ +  ab^ +  b - \frac{\chi}{{\sqrt S }}\left( {a^ +  a^ +  b + b^ +  aa} \right) \\
 \end{array}
\end{eqnarray}
is the condensate part of the Hamiltonian with $\mu _b  = 2\mu  + u_{bb} /S$ the modified chemical potential of PPs, and where $\delta H = H\left( {a_{\bf{k}} ,b_{\bf{k}} } \right)$ is the perturbation part and it is a complex function of the non-condensate modes. Here, we mention that at the condensate state we only need consider the case of single mode coupling, thus we can treat the coupling matrix element $\chi$ as an adjustable constant. Note that in (11) we also drop the zero-momentum subscript of the creation and annihilation operators of photons and PPs. In the present work, we mainly aim to investigate the phenomenon of BEC of photons and PPs, thus, hereafter we will ignore the perturbation part and approximately write the Hamiltonian as the form $H_\mu   \approx H_0$. The Hamiltonian commutes with the total photon number $N = a^ +  a + 2b^ +  b$ and $n = N/S$ is the particle density.

Up to now we have not made a careful distinction between the two possible condensate phases. However, for the mixed system, working out the ground-state phase diagram is very important. Here, we intend to employ the variational principle method for finding the ground-state configurations of the present system and examining their dependence from the microscopic parameters. In other word, we aim to work out the ground-state phase diagram of the mixed system starting from the study the semiclassical equation. Clearly, we know that the chemical potential $\mu$ of the mixed system allows the total photon number (whether free or bound into PPs) to fluctuate around some constant average value $N$, then the total number of photons need only be conserved on the average value. For convenience, hereafter we assume that $N$ is an even number and thus $M = N/2$ denotes the maximum number of the PPs. Furthermore, in this case we also introduce a new operation, namely the double photon creation operation with the relation
\begin{eqnarray}
\begin{array}{l}
 \left( {c^ +  } \right)^m \left| 0 \right\rangle  \equiv \sqrt {\frac{{m!}}{{\left( {2m} \right)!}}} \left( {a^ +  } \right)^{2m} \left| 0 \right\rangle  \\
 {\kern 1pt} {\kern 1pt} {\kern 1pt} {\kern 1pt} {\kern 1pt} {\kern 1pt} {\kern 1pt} {\kern 1pt} {\kern 1pt} {\kern 1pt} {\kern 1pt} {\kern 1pt} c^ +  c\left| 0 \right\rangle  \equiv 2a^ +  a\left| 0 \right\rangle  \\
 \end{array},
\end{eqnarray}
where $\left| 0 \right\rangle$ is the vacuum state. Using the new operator, we construct the Gross-Pitaevskii (GP) states [28]
\begin{eqnarray}
\left| {\psi _{{\rm{GP}}} } \right\rangle  = \frac{1}{{\sqrt {M!} }}\left[ {\alpha c^ +   + \beta b^ +  } \right]^M \left| 0 \right\rangle
\end{eqnarray}
as the trial macroscopic state. Here, $\alpha  = \left| \alpha  \right|e^{i\theta _a }$ and $\beta  = \left| \beta  \right|e^{i\theta _b }$ are complex amplitudes with $\left| \alpha  \right|^2  = N_a /N$ and $\left| \beta  \right|^2  = 2N_b /N$ the photon and PP densities, respectively. $\theta _a$ and $\theta _b$ (real valued) denoted the phases of each species. Obviously, the parameters $\alpha$ and $\beta$ satisfy the normalized condition $\left| \alpha  \right|^2  + \left| \beta  \right|^2  = 1$. With the help of the GP state, the semiclassical model Hamiltonian $\bar H(\alpha ,\beta )$ is given by
\begin{eqnarray}
\begin{array}{l}
 \bar H = \mathop {\lim }\limits_{N \to \infty } \frac{{\left\langle {\psi _{GP} } \right|H_\mu  \left| {\psi _{GP} } \right\rangle }}{{M{\chi}\sqrt {2n} }} \\
 {\kern 1pt} {\kern 1pt} {\kern 1pt} {\kern 1pt} {\kern 1pt} {\kern 1pt} {\kern 1pt} {\kern 1pt} {\kern 1pt}  = \left[ { - \mu _b \left| \beta  \right|^2  + 2u_{aa} n\left| \alpha  \right|^4  + u_{bb} n\left| \beta  \right|^4 } \right./2 \\
 {\kern 1pt} {\kern 1pt} {\kern 1pt} {\kern 1pt} {\kern 1pt} {\kern 1pt} {\kern 1pt} {\kern 1pt} {\kern 1pt} {\kern 1pt} {{\left. { + u_{ab} n\left| \alpha  \right|^2 \left| \beta  \right|^2 } \right]} \mathord{\left/
 {\vphantom {{\left. { + u_{ab} n\left| \alpha  \right|^2 \left| \beta  \right|^2 } \right]} {{\chi}\sqrt {2n} }}} \right.
 \kern-\nulldelimiterspace} {{\chi}\sqrt {2n} }} - 2\left| \alpha  \right|^2 \sqrt {\left| \beta  \right|^2 } \cos \theta  \\
 \end{array},
\end{eqnarray}
where $\theta  = \theta _b  - 2\theta _a$ is the phase difference. Considering the conserved condition $\left| \alpha  \right|^2  + \left| \beta  \right|^2  = 1$, we next introduce a new variables $s = \left| \alpha  \right|^2  - \left| \beta  \right|^2$. Using the new notation, we rewrite the model Hamiltonian as
\begin{eqnarray}
\bar H =  - \lambda s^2  - 2\gamma s + \xi  - \sqrt {2\left( {1 - s} \right)} (1 + s)\cos \theta,
\end{eqnarray}
with

$\begin{array}{l}
 \lambda  = \frac{{\sqrt {2n} }}{{\chi}}\left( {\frac{{u_{ab} }}{4} - \frac{{u_{aa} }}{2} - \frac{{u_{bb} }}{8}} \right) \\
 \gamma  = \frac{{\sqrt {2n} }}{{\chi}}\left( {\frac{{u_{bb} }}{8} - \frac{{u_{aa} }}{2} - \frac{{\mu _b }}{{4n}}} \right) \\
 \xi  = \frac{{\sqrt {2n} }}{{\chi}}\left( {\frac{{u_{aa} }}{2} + \frac{{u_{ab} }}{4} + \frac{{u_{bb} }}{8} - \frac{{\mu _b }}{n} } \right) \\
 \end{array}$.

\hspace{-20.pt}According to the variational principle, we minimize the energy $\bar H(\alpha ,\beta )$ with $s$ and $\theta$ as variational parameters. We then obtain the optimum values [i.e.($\bar s$, $\bar \theta$)] of parameters for the ground state as follows:
\begin{eqnarray}
\left( {\bar s,\bar \theta } \right) = \left\{ \begin{array}{l}
 ( - 1,{\kern 1pt} {\kern 1pt} {\kern 1pt} {\kern 1pt} {\kern 1pt} {\kern 1pt} {\kern 1pt} {\kern 1pt} {\kern 1pt} \theta ),{\kern 1pt} {\kern 1pt} {\kern 1pt} {\kern 1pt} {\kern 1pt} {\kern 1pt} {\kern 1pt} {\kern 1pt} {\kern 1pt} {\kern 1pt} {\kern 1pt} {\kern 1pt} {\kern 1pt} {\kern 1pt} {\kern 1pt} {\kern 1pt} {\kern 1pt} {\kern 1pt} {\kern 1pt} {\kern 1pt} {\kern 1pt} {\kern 1pt} {\kern 1pt} {\kern 1pt} {\kern 1pt} {\kern 1pt} {\kern 1pt} {\kern 1pt} {\kern 1pt} {\kern 1pt} {\kern 1pt} {\kern 1pt} {\kern 1pt} {\kern 1pt} {\kern 1pt} {\kern 1pt} {\kern 1pt} {\kern 1pt} {\kern 1pt} {\kern 1pt} {\kern 1pt} {\kern 1pt} {\kern 1pt} {\kern 1pt} {\kern 1pt} \gamma  - \lambda  + 1 < 0 \\
 (s,{\kern 1pt} {\kern 1pt} {\kern 1pt} {\kern 1pt} 0{\kern 1pt} {\kern 1pt} {\kern 1pt} or{\kern 1pt} {\kern 1pt} {\kern 1pt} \pi ),{\kern 1pt} {\kern 1pt} {\kern 1pt} {\kern 1pt} {\kern 1pt} {\kern 1pt} {\kern 1pt} {\kern 1pt} {\kern 1pt} {\kern 1pt} {\kern 1pt} {\kern 1pt} {\kern 1pt} {\kern 1pt} {\kern 1pt} {\kern 1pt} {\kern 1pt} {\kern 1pt} {\kern 1pt} {\kern 1pt} {\kern 1pt} {\kern 1pt} {\kern 1pt} {\kern 1pt} {\kern 1pt} {\kern 1pt} {\kern 1pt} {\kern 1pt} {\kern 1pt} {\kern 1pt} {\kern 1pt} {\kern 1pt} {\kern 1pt} {\kern 1pt} {\kern 1pt} {\kern 1pt} \gamma  - \lambda  + 1 > 0 \\
 \end{array} \right.,
\end{eqnarray}
where $- 1 < s < 1$ is the solution of the equation $\lambda s + \gamma  = \left( {3s - 1} \right)/2\sqrt {2(1 - s)}$ (the explicit value can be obtain by graphical solution method, it is generally too messy to be shown here). The result, together with the fact that the parameters $\left| \alpha  \right|^2$ and $\left| \beta  \right|^2$ denote the photon and PP densities, indicates that when $\gamma  - \lambda  + 1 < 0$ the system converts from the mixed photon-PP phase to the pure PP phase. We therefore can interpret this line $\gamma  - \lambda  + 1 = 0$ as the threshold coupling for the formation of a predominantly PP state. Here, it is to be mentioned that for the pure PP phase, $\bar s =  - 1$ thus the relative phase $\theta$ cannot be defined.

\section{Entanglement between photons and photon pairs}
\subsection{Entanglement of the ground state}
At the BEC state, one may consider the mixed gas of photons and PPs as a bipartite system of two modes. For the present system, the entanglement of two modes is always closely associated with the phase transition of the system. Moreover, the two modes, be they spatially separated, and differing in some internal quantum number, are clearly distinguishable subsystems. Thus, the state of each mode can be characterized by its occupation number. By using the fact that the total number of photons $N$ is constant, a general state of the system (in the Heisenberg picture) can be written for even $N$ in terms of the Fock states by
\begin{eqnarray}
\left| \psi  \right\rangle  = \sum\limits_{m = 0}^M {c_m } \left| {2m,M - m} \right\rangle,
\end{eqnarray}
where $m$ is the half population of particles in photon mode $a$, and $c_m$ is the coefficients of the state. In the Fock representation, the GP state also can be reexpressed as $\left| {\psi _{GP} } \right\rangle  = \sum\limits_{m = 0}^M {g_m } \left| {2m,M - m} \right\rangle$, with coefficients $g_m  = \sqrt {\frac{{M!}}{{m!\left( {M - m} \right)!}}} \alpha ^m \beta ^{M - m}$. The standard measure of entanglement of the bipartite system is the entropy of entanglement $S(\rho )$
\begin{eqnarray}
S(\rho ) =  - \sum\limits_{m = 0}^M {\left| {c_m } \right|^2 } \log _2 \left( {\left| {c_m } \right|^2 } \right),
\end{eqnarray}
which is the von Neumann entropy of the reduced density operator of either of the subsystems[29]. In the present system, the maximal entanglement also can be obtain by optimizing the expression (18) with respect to $\left| {c_m } \right|^2$. By imposing the normalization condition $\sum\limits_{m = 0}^M {\left| {c_m } \right|^2 }  = 1$, we finally get $S_{\max }  = \log _2 \left( {M + 1} \right)$, which is related to the dimension $M + 1$ of the Hilbert space of the individual modes.

Using expression (18) and the coefficients $c_m$ obtained through exact diagonalization of the Hamiltonian (11) as done in the atom-molecule model [30], we plot in Fig. 2 the entropy of entanglement of the ground state as a function of the parameters $\lambda$ and $\gamma$. We note that in this letter we restrict our attentions to the repulsive case, i.e., we restrict $\lambda  \le 0$ throughout the letter. From Fig. 2, we observe that the entanglement entropy exhibits a sudden decrease close $\gamma  - \lambda  + 1 = 0$. This is indicative of the fact that across the line $\gamma  - \lambda  + 1 = 0$ a quantum phase transition occurs.
\begin{figure}[t]
  \includegraphics[width=8cm,]{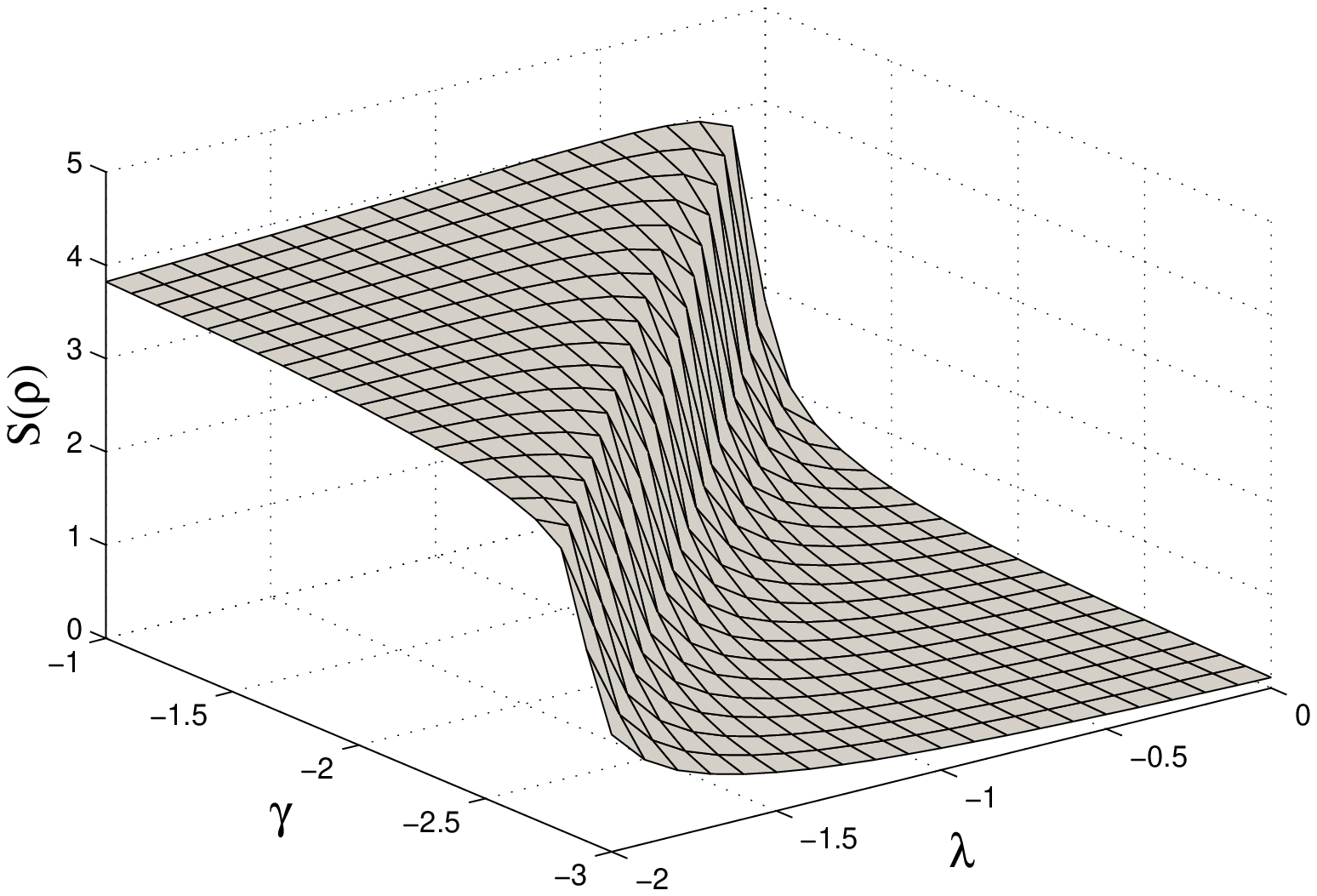}
  \caption{Variation in the entropy of entanglement of the ground state with respect
to the parameters $\lambda$ and $\gamma$. Here, we set $N=200$ and $u_{aa}  = u_{bb}/4  = 0.25$. }
\end{figure}

To gain more information associated with the quantum phase transition of the present system, we also depict in Fig. 3 the entropy of entanglement (solid line) and the expectation value (dashed line) of the scaled PP number operator of the ground state as a function $\gamma$ for fixed parameter value $\lambda$. From Fig. 3, we see that the average value of the number of PPs increases as $\gamma$ increases. Especially, when $\gamma  - \lambda  + 1 < 0$, the average number of PPs is maximal. The result confirms that there indeed exists a phase transition for the present system in the ground state. Furthermore, we also find that the ground-state entanglement entropy is not maximal at the critical line, i.e. in the region $\gamma  - \lambda  + 1 > 0$, the system is always strongly entangled. Ref. (28) gives the property responsible for the long-range correlation. Additionally, we also consider that the trait is associated with the symmetry-broking of the coupling term of the system. Due to the asymmetric form of the coupling term, the pure photon condensation will be forbidden. As a result, in the mixed condensate phase the imbalance $\left( {2N_b  - N_a } \right)/N$ between the two modes is always very small, which is responsible for the strongly entanglement.
\begin{figure}[t]
  \includegraphics[width=9cm,]{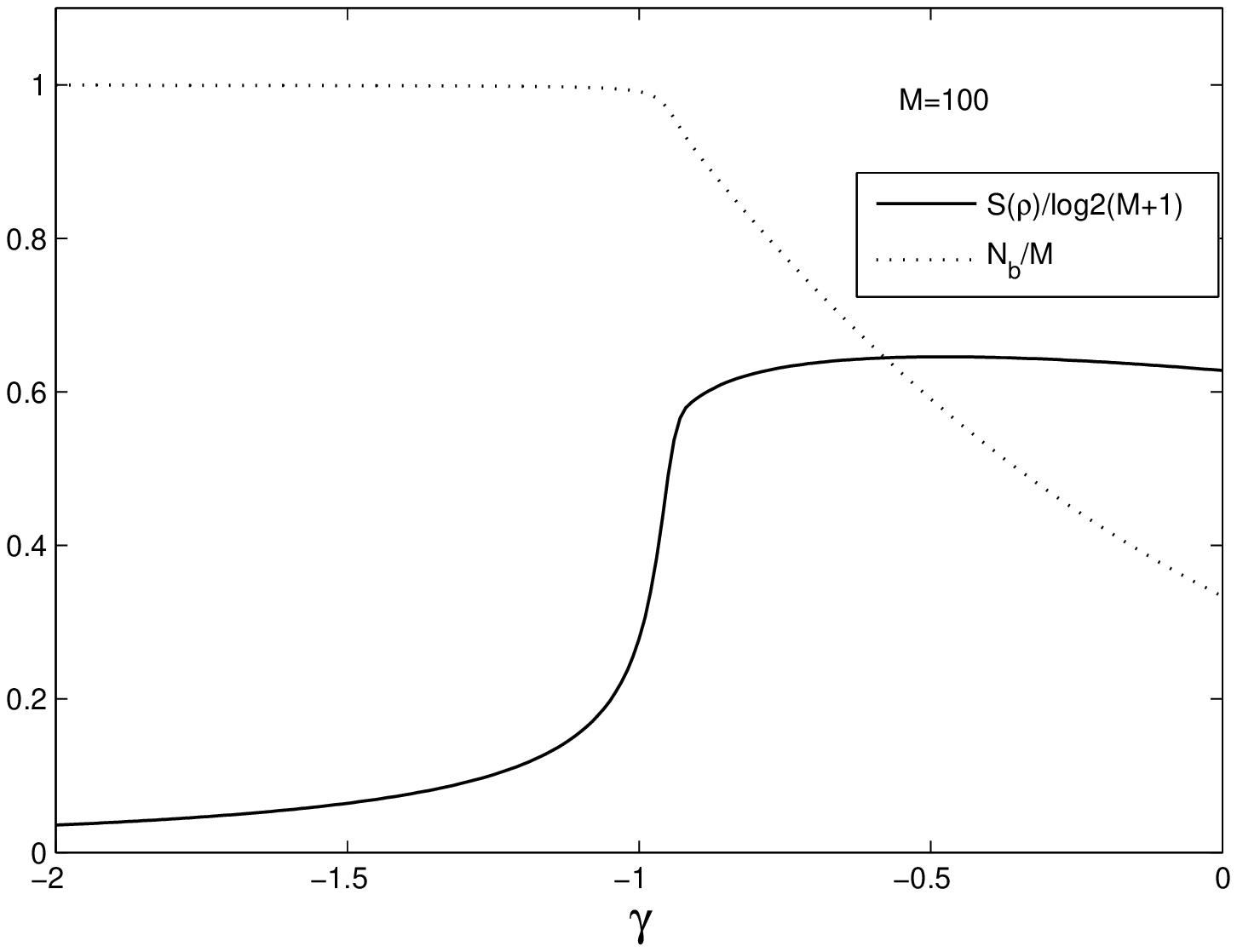}
  \caption{The average photon pair occupation number and the entanglement entropy for the ground state as a function of the $\gamma$ for fixed parameter value $\lambda=0$. Here, we set $N=200$ and $u_{aa}  = u_{bb}/4 = 0.25$. }
\end{figure}

In addition, if we connect the non-vanishing photon mass $m_{ph}$ to the longitudinal wave number $k_z$ by the relation $m_{{\rm{ph}}}  = \hbar k_z /c$ with $k_z  = q\pi /D_0$, then we find that the quantum phase transition of the photon system can be interpreted as second harmonic generation. When $\gamma  - \lambda  + 1 < 0$, almost all photons with frequency $\omega  = ck_z$ couple into PPs with frequency $\omega  = 2ck_z$. In this case, the entanglement between the photons and PPs is very small, and the entropy of entanglement is close to zero.

\subsection{Dynamics of entanglement}

In the above analysis, we have investigated the entanglement of the ground state. We found that in the ground state, across the phase transition line the entanglement entropy exhibits a sudden change. To gain a better understanding of the influence of ground-state phase transition to entanglement, in this subsection we investigate the dynamics of entanglement. In studying the dynamics of the system, we first need express a general state in the form of temporal evolution (i.e., need change the expression of a general state from Heisenberg picture to Schrodinger picture). Following the standard procedure, we can obtain
\begin{eqnarray}
\begin{array}{l}
 \left| {\psi \left( t \right)} \right\rangle  = U\left( t \right)\left| {\psi \left( 0 \right)} \right\rangle  \\
  = \sum\limits_{m = 0}^M {c_m } \left( t \right)\left| {2m,M - m} \right\rangle  \\
 \end{array},
\end{eqnarray}
where, $U\left( t \right) = \sum\limits_{n = 0}^M {\left| {\psi _n } \right\rangle \left\langle {\psi _n } \right|} \exp \left( { - iE_n t} \right)$ is the temporal operator with $\left| {\psi _n } \right\rangle$ the eigenstates of the system having energy $E_n$, and $\left| {\psi \left( 0 \right)} \right\rangle$ is the initial state. Here, the time dependence of coefficients  $c_m \left( t \right)$ are given by $c_m \left( t \right) = \left\langle {2m,M - m} \right|U\left( t \right)\left| {\psi \left( 0 \right)} \right\rangle$. Subsequently, the entanglement entropy given in (18) can be rewritten as $S(\rho ) =  - \sum\limits_{m = 0}^M {\left| {c_m \left( t \right)} \right|^2 } \log _2 \left( {\left| {c_m \left( t \right)} \right|^2 } \right)$. In this case, the entanglement entropy depends on both the choice of initial states and the value of microscopic parameters $\left\{ {\lambda ,\gamma } \right\}$. At the present work, we consider that the mixed system is in the BEC state, thus here choosing GP state as the initial state is suitable. By adjusting the GP coefficients $\left\{ {\left| \alpha  \right|^2 ,\left| \beta  \right|^2 } \right\}$ and the microscopic parameters $\left\{ {\lambda ,\gamma } \right\}$, in this subsection we also want to know if the ground-state phase transition also characterizes different dynamics.

The time evolution of the entanglement entropy for different initial state and interaction parameters is shown in Fig. 4. From Fig.4 we observe the features of quantum dynamics, such as the collapse and revival of oscillations and non-periodic oscillations. Additionally, we also find that the amplitude of the entanglement entropy is smaller in the region $\gamma  - \lambda  + 1 < 0$ contrast with in the region $\gamma  - \lambda  + 1 > 0$. Especially, we note that the greater the imbalance $\left| \beta  \right|^2  - \left| \alpha  \right|^2$ between the two modes in the initial state, the clearer the difference can be observed.
\begin{figure}[t]
  \includegraphics[width=9cm,]{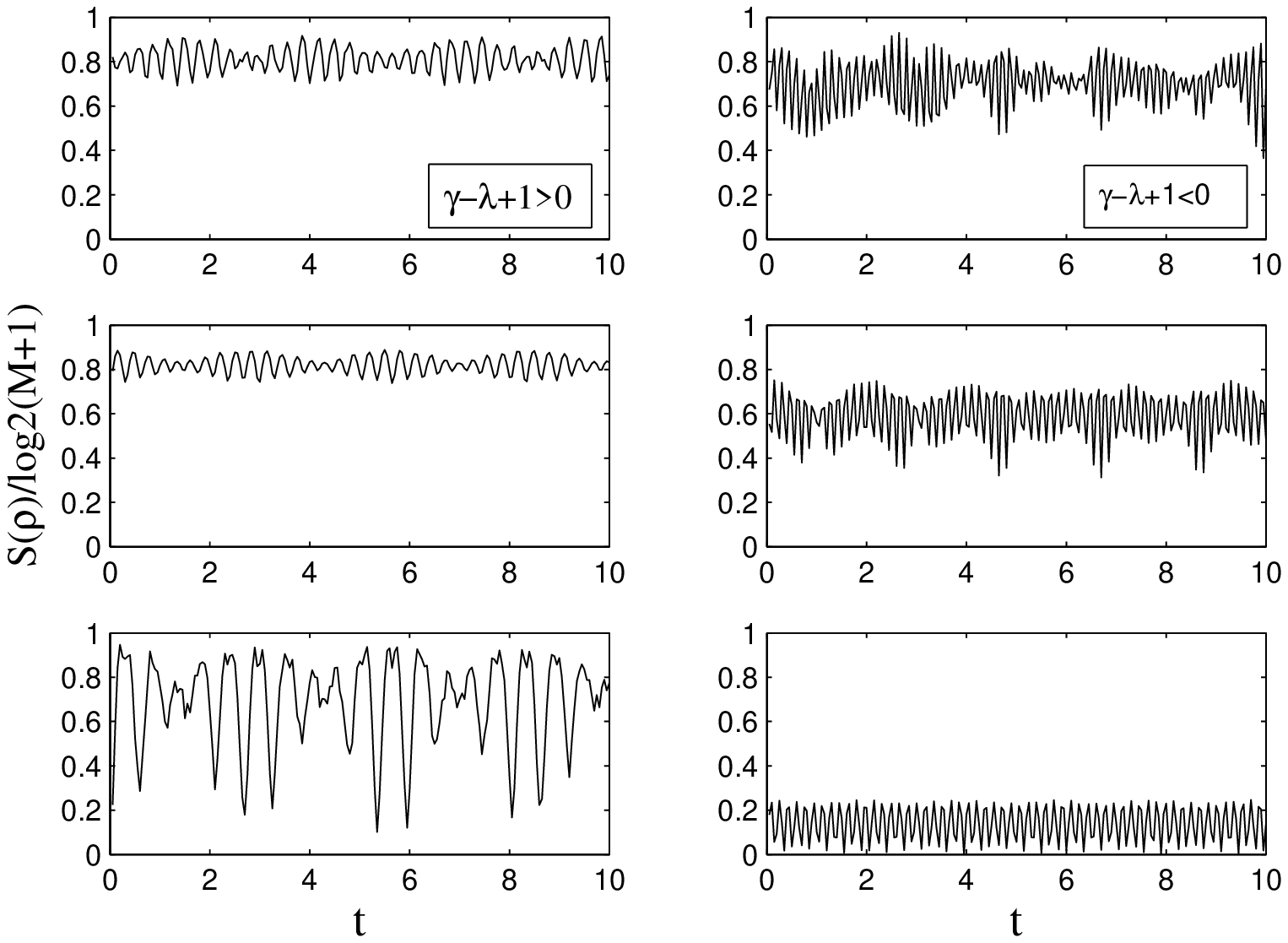}
  \caption{Time evolution of the entanglement entropy for different initial states $\left| {\psi \left( {\left| \alpha  \right|^2 ,\left| \beta  \right|^2 } \right)} \right\rangle$ and microscopic parameters $\left\{ {\lambda ,\gamma } \right\}$. Form top to bottom the GP coefficients used are $\left\{ {\left| \alpha  \right|^2 ,\left| \beta  \right|^2 } \right\} = \left\{ {0.5,0.5} \right\}$, $\left\{ {\left| \alpha  \right|^2 ,\left| \beta  \right|^2 } \right\} = \left\{ {0.25,0.75} \right\}$ and $\left\{ {\left| \alpha  \right|^2 ,\left| \beta  \right|^2 } \right\} = \left\{ {0,1} \right\}$. From left to right the microscopic parameters used are $\left\{ {\gamma ,\lambda } \right\} = \left\{ { - 2,0} \right\}$ and $\left\{ {\gamma ,\lambda } \right\} = \left\{ { - 0.5,0} \right\}$, respectively. Here, we set $N=20$ and $u_{aa}  = u_{bb}/4  = 0.25$.
 }
\end{figure}
To understand the physical reason for the above phenomenon, we also need rewrite the general state given in (19) in terms of the eigenstates of the system $\left| {\psi \left( t \right)} \right\rangle  = \sum\limits_{n = 0}^M {c(n,t)} \left| {\psi _n } \right\rangle$, where
$\left| {c(n,t)} \right|^2 =\left| {c(n,0)} \right|^2 = \left| {\left\langle {\psi _n } \right|\exp ( - iE_n t)\left| {\psi \left( 0 \right)} \right\rangle } \right|^2$ can be explained as the transition probability of the system from the initial state $\left| {\psi \left( 0 \right)} \right\rangle$ to the corresponding energy eigenstates $\left| {\psi _n } \right\rangle$ at any time $t$. We have already known that for the ground state across the phase transition line $\gamma  - \lambda  + 1 = 0$ the entanglement entropy exhibits a sudden decrease. This is why in the region $\gamma  - \lambda  + 1 < 0$ the amplitude of the entanglement entropy becomes smaller. In addition, in this phase transition region we also investigative the dependence relation between the ground-state transition probability $\left| {c\left( {0,t} \right)} \right|^2$ and the imbalance $\left| \beta  \right|^2  - \left| \alpha  \right|^2$ of the initial state. The result is shown in Fig. 5. From Fig. 5, it is obvious that with the increasing of the initial-state imbalance the ground-state transition probability becomes greater. Thus the greater imbalance between the two modes in the initial state can lead the clearer difference of the amplitude for different region.
\begin{figure}[t]
  \includegraphics[width=9cm,]{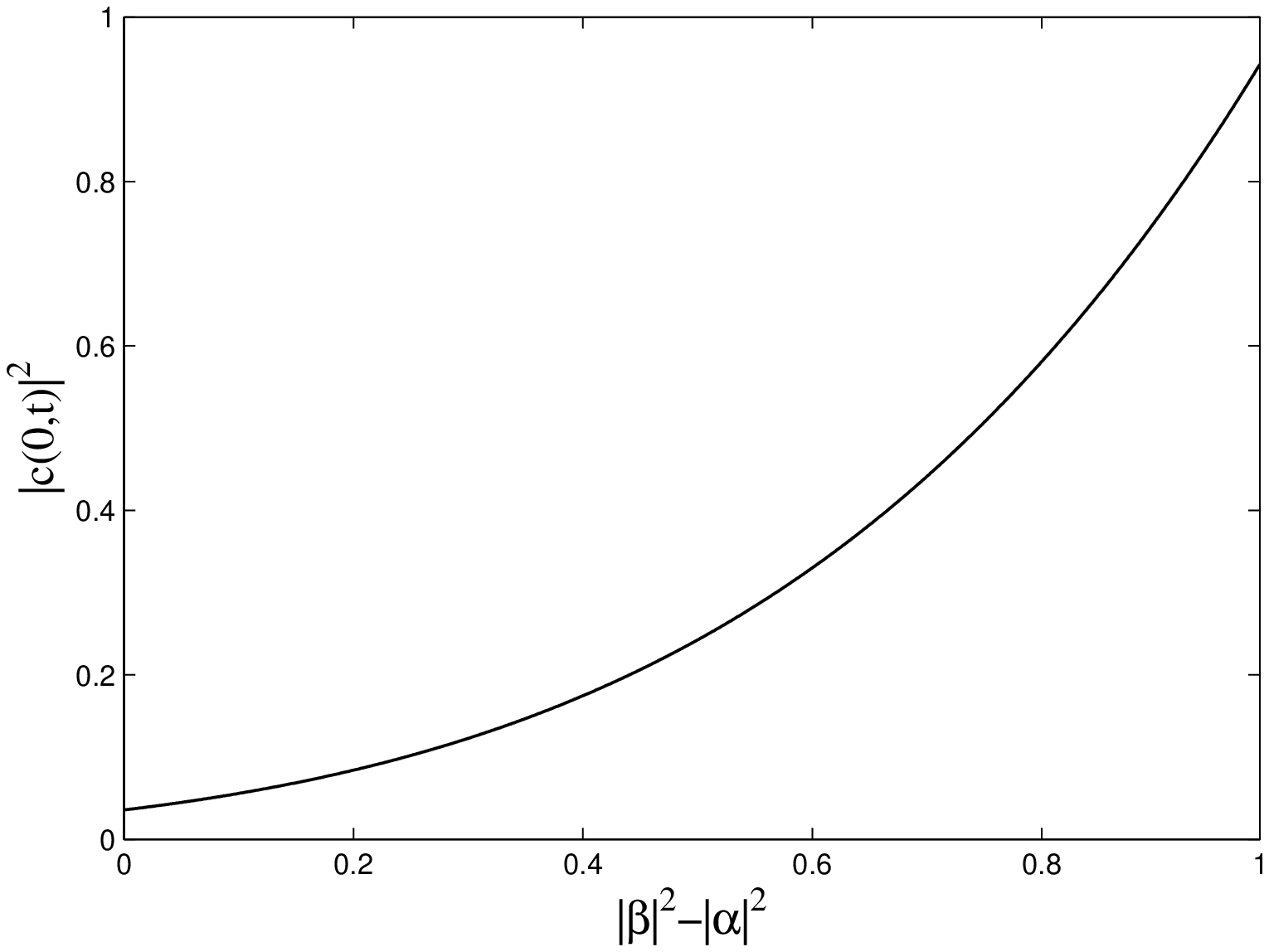}
  \caption{Variation of the ground-state transition probability $\left| {c\left( {0,t} \right)} \right|^2$ with respect to the initial-state imbalance $\left| \beta  \right|^2  - \left| \alpha  \right|^2$. Here, we set $N=20$ and $u_{aa}  = u_{bb}/4  = 0.25$.}
\end{figure}

\section{Conclusion}

In this work, we have proposed a 2D model consisting of photons and PPs. In the model, the mixed gas of photons and PPs is formally equivalent to a 2D system of massive bosons with non-vanishing chemical potential, which implies the existence of two possible condensate phase. Based on the GP state and using the variational method, we have also discussed the quantum phase transition of the mixed gas and have obtained the critical coupling line analytically. Especially, we have found that the phase transition of the photon gas can be interpreted as second harmonic generation. Moreover, by investigating the entanglement entropy in the ground state and general state, we have illustrated how the entanglement between photons and PPs can be associated with the phase transition of the system.
\label{}






\begin{thebibliography}{00}
\bibitem{1}M. H. Anderson, J. R. Ensher, M. R. Matthews, C. E.
Wieman, and E. A. Cornell, Science \textbf{269} (1995) 198.
\bibitem{2}K. B. Davis, M-O. Mewes, M. R. Andrews, N. J. van Druten,
D. M. Kurn, and W. Ketterle, Phys. Rev. Lett. \textbf{75} (1995) 3969.
\bibitem{3}C. C. Bradley, C. A. Sackett, and R. G. Hulet, Phys. Rev.
Lett. \textbf{78} (1997) 985.
\bibitem{4}S. Jochim, \emph{et al}, Science \textbf{302} (2003) 2101.
\bibitem{5}H. Deng, G. Weihs, J. Bloch, and Y. Yamamoto, Science \textbf{298} (2002) 199.
\bibitem{6}J. Kasprzak, \emph{et al}, Nature \textbf{443} (2006) 409.
\bibitem{7}R. Balili, V. Hartwell, D. Snoke, L. Pfeiffer, and K.
West, Science \textbf{316} (2007) 1007.
\bibitem{8}S.O. Demokritov, \emph{et al}, Nature \textbf{443} (2006) 430.
\bibitem{9}G. Ortiz and J. Dukelsky, Phys. Rev. A \textbf{72} (2005) 043611.
\bibitem{10}J. Klaers, J. Schmitt, F. Vewinger, and M. Weitz,
Nature \textbf{468} (2010) 545.
\bibitem{11}J. Klaers, F. Vewinger, and M. Weitz, Nature Phys. \textbf{6} (2010) 512.
\bibitem{12}D. F. Walls, and P. Zoller Phys. Rev. Lett. \textbf{47} (1981) 709.
\bibitem{13}G. Rempe, F. Schmidt-Kaler, and H. Walther, Phys. Rev. Lett \textbf{64} (1990) 2783.
\bibitem{14}O. V. Kibis, G. Ya. Slepyan, S. A. Maksimenko, and A. Hoffmann, Phys. Rev. Lett \textbf{102} (2009) 023601.
\bibitem{15}D. F. Walls, and R. Barakat, Phys. Rev. A \textbf{1} (1970) 446.
\bibitem{16}Z. Cheng, Phys. Rev. Lett. \textbf{67} (1991) 2788.
\bibitem{17}T. Yamamoto, M. Koashi, S. K. \"{O}zdemir, and N. Imoto, Nature \textbf{421} (2003) 343.
\bibitem{18}Z. D. Walton, A. V. Sergienko, B. E. A. Saleh, and M. C. Teich, Phys. Rev. A \textbf{70} (2004) 052317.
\bibitem{19}W. Denk, J. H. Strickler, and W. W. Webb, Science \textbf{6} (1990) 73.
\bibitem{20}T. Binoth, J.Ph. Guillet, E. Pilon, and M. Werlen, Eur. Phys. J. C \textbf{16} (2000) 311.
\bibitem{21}Z. Cheng, J. Opt. Soc. Am. B \textbf{19} (2002) 1962.
\bibitem{22}N. Bloembergen, Non-Linear Optics (W. A. Benjamin, Inc., New York, 1965)
\bibitem{23}J. Klaers, J. Schmitt, T. Damm, F. Vewinger, and M. Weitz Appl. Phys. B \textbf{105} (2011) 17.
\bibitem{24}It may be asked why this term can be ignored. The answer is connected with the fact that in this letter we only consider the infinite particle number case, i.e., we consider the total photon number $N$ as a constant. In the case, the chemical potential term $\mu N_a$ cannot give new restriction to the Hamiltonian. Especially, in the following analysis we will point out that at the BEC state the mixed gas of photons and PPs can convert from the mixed photon-PP condensate phase to the pure PP condensate phase. In the pure PP condensate phase, the term $\mu N_a$ is meaningless. We therefore ignore the restriction term $\mu N_a$ in the model Hamiltonian.
\bibitem{25}L. Radzihovsky, J. Park, and P. B. Weichman, Phys. Rev. Lett. \textbf{92} (2004) 160402.
\bibitem{26}M. W. J. Romans, R. A. Duine, S. Sachdev, and H. T. C. Stoof, Phys. Rev. Lett. \textbf{93} (2004) 020405;
\bibitem{27}G. Santos, A. Tonel, A. Foerster, and J. Links, Phys. Rev. A \textbf{73} (2006) 023609; M. Duncan, A. Foerster, J. Links, E. Mattei, N. Oelkers, and A. P. Tonel, Nucl. Phys. B \textbf{767} (2007) 227; G. Santos, A. Tonel, A. Foerster, and J. Links, Phys. Rev. A \textbf{73} (2006) 023609.
\bibitem{28}S. C. Li, J. Liu, and L. B. Fu, Phys. Rev. A \textbf{83}, (2011) 042107; S. C. Li, and L. B. Fu, Phys. Rev. A \textbf{84}, (2011) 023605.
\bibitem{29}A. P. Hines, R. H. McKenzie, and G. J. Milburn, Phys. Rev. A \textbf{67}, (2003) 013609.
\bibitem{30}A. P. Tonel, J. Links, and A. Foerster, J. Phys. A:Math. Gen. \textbf{38}, (2005) 1235; J. Links, H. Q. Zhou, R. H. McKenzie, and M. D. Gould, J. Phys. A \textbf{36} (2003) R63.

\end{thebibliography}

\end{document}